\documentclass[final]{aipproc}
\usepackage{amsmath, graphicx}
\layoutstyle{6x9}

\newcommand{\llangle}{\langle\!\langle}
\newcommand{\rrangle}{\rangle\!\rangle}

\begin{document}

\title{\mbox{Noise and Bistabilities in Quantum Shuttles}}


\author{Christian Flindt}{
  address={NanoDTU, MIC -- Department of Micro and Nanotechnology, Technical University of Denmark, \mbox{DTU - Building 345 East}, \mbox{DK - 2800 Kongens Lyngby},
  Denmark}
}

\author{Tom\'{a}\v{s} Novotn\'{y}}{
  address={Nano-Science Center, University of Copenhagen - Universitetsparken 5, \mbox{DK - 2100 Copenhagen \O},
  Denmark},
  altaddress={Department of Electronic Structures, Faculty of
  Mathematics and Physics, \mbox{Charles University - Ke Karlovu 5, 121
  16 Prague}, Czech Republic}
}

\author{Antti-Pekka Jauho}{
  address={NanoDTU, MIC -- Department of Micro and Nanotechnology, Technical
  University of Denmark, \mbox{DTU - Building 345 East}, \mbox{DK -
  2800 Kongens Lyngby}, Denmark}
}

\begin{abstract}
We present a study of current fluctuations in two models proposed
as quantum shuttles. Based on a numerical evaluation of the first
three cumulants of the full counting statistics we have recently
shown that a giant enhancement of the zero-frequency current noise
in a single-dot quantum shuttle can be explained in terms of a
bistable switching between two current channels. By applying the
same method to a quantum shuttle consisting of a vibrating quantum
dot array, we show that the same mechanism is responsible for a
giant enhancement of the noise in this model, although arising
from very different physics. The interpretation is further
supported by a numerical evaluation of the finite-frequency noise.
For both models we give numerical results for the effective
switching rates.
\end{abstract}

\maketitle

\emph{Introduction} -- In 1998 Gorelik \emph{et al.} proposed a
nano-electromechanical system (NEMS), the charge shuttle,
consisting of a movable nanoscopic grain coupled via tunnel
barriers to source and drain electrodes \cite{gorelik_1998}.
Originally the motion of the grain was modelled using a classical
harmonic oscillator. Here we present a study of current
fluctuations in two models of (quantum) shuttles, where the
oscillator is quantized.

\emph{Models} -- Two models have been proposed as quantum shuttles
(the 1-dot shuttle \cite{novotny_2003} and the 3-dot shuttle
\cite{armour_2002}). The 1-dot shuttle consists of a single
mechanically oscillating quantum dot situated between two leads.
In the 3-dot shuttle the mechanically oscillating quantum dot is
flanked by two static dots, thus making up an array of dots. Both
devices are operated in the strong Coulomb blockade regime, and
consequently only one excess electron at a time is allowed in the
device. In the 1-dot (3-dot) model the coupling to the leads (the
interdot coupling) depends exponentially on the position of the
vibrating dot. For detailed descriptions of the models we refer to
Refs. \cite{novotny_2003,armour_2002,flindt_2004}.

Both models are described using the language of quantum
dissipative systems \cite{weiss}. As the ``system'' we take in the
1-dot model (3-dot model) the single (three) electronic state(s)
of the occupied dot (array) and the unoccupied state plus the
quantum harmonic oscillator with natural frequency $\omega_0$. In
the limit of a high bias between the leads \cite{gurvitz_1996},
and assuming that the oscillator is damped due to a weak coupling
to a heat bath, the time evolution of the reduced density matrix
of the system $\hat{\rho}(t)$ is governed by a Markovian
generalized master equation (GME) of the form
\cite{novotny_2003,armour_2002,flindt_2004}
\begin{equation}
\dot{\hat{\rho}}(t)=\mathcal{L}\hat{\rho}(t)=(\mathcal{L}_{\mathrm{coh}}
+\mathcal{L}_{\mathrm{damp}}+\mathcal{L}_{\mathrm{driv}})\hat{\rho}(t).
\label{eq_genericGME}
\end{equation}
Here $\mathcal{L}_{\mathrm{coh}}$ describes the internal coherent
dynamics of the system, while $\mathcal{L}_{\mathrm{damp}}$ and
$\mathcal{L}_{\mathrm{driv}}$ give the damping and the coupling to
the leads, respectively. In the following we consider the
stationary state defined by
$\dot{\hat{\rho}}^{\mathrm{stat}}(t)=\mathcal{L}\hat{\rho}^{\mathrm{stat}}(t)=0$.
The GME is only valid in the high-bias limit, and hence we cannot
use the applied bias as a control parameter. Instead, we vary in
the 1-dot model the strength of the damping, denoted $\gamma$, and
in the 3-dot model the difference between the energy levels
corresponding to the outer dots, referred to as the \emph{device
bias} and denoted $\varepsilon_b$.

\begin{figure}
  \includegraphics[height=.16\textheight, trim = 0 0 0 0, clip]{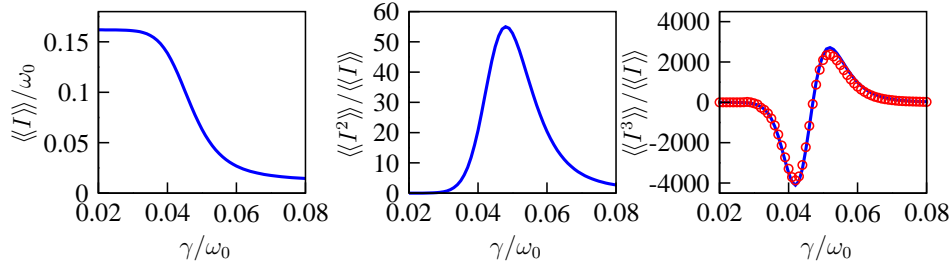}
  \caption{First three cumulants for the 1-dot model as a function of the damping strength $\gamma$ (model parameters correspond to Fig. 3 in Ref. \cite{flindt_2005}). The shuttling channel current
  is $I_{\mathrm{shut}}=\omega_0/2\pi$ and the tunneling channel current $I_{\mathrm{tun}}=0.0082\omega_0$ ($e=1$). Full lines
  indicate numerical results, while circles show the \mbox{(semi-)} analytic results for the third cumulant.
  The central panel shows the giant enhancement of the zero-frequency noise. (Reproduced from Ref. \cite{flindt_2005}).}
  \label{fig_1dotFCS}
\end{figure}

\emph{Theory} -- We have recently developed a systematic theory
for the calculation of the $n$'th cumulant of the current
$\llangle I^n\rrangle $ for NEMS described by a Markovian GME of
the form given in Eq. (\ref{eq_genericGME}) \cite{flindt_2005}. In
Ref. \cite{flindt_2005} a numerical evaluation of the first three
cumulants showed that the 1-dot model in a certain parameter
regime behaves as a bistable system switching slowly between two
current channels. The first three cumulants of a bistable system
switching slowly (compared to the electron transfer rates) with
rates $\Gamma_{1\leftarrow2}$ and $\Gamma_{2\leftarrow1}$ between
two current channels 1 and 2 with corresponding currents $I_1$ and
$I_2$, respectively, are \cite{jordan_2004}
\begin{equation}
\begin{split}
\llangle I\rrangle &=\frac{I_{1}\Gamma_{1\leftarrow2}
+I_{2}\Gamma_{2\leftarrow1}}
{\Gamma_{2\leftarrow1}+\Gamma_{1\leftarrow2}}\ ,
\\ \llangle I^2\rrangle
&=2(I_{1}-I_{2})^2\frac{\Gamma_{1\leftarrow2}\Gamma_{2\leftarrow1}}{(\Gamma_{1\leftarrow2}
+\Gamma_{2\leftarrow1})^3}\ ,
\\ \llangle I^3\rrangle&=
6(I_{1}-I_{2})^3\frac{\Gamma_{1\leftarrow2}\Gamma_{2\leftarrow1}
(\Gamma_{2\leftarrow1}-\Gamma_{1\leftarrow2})}
{(\Gamma_{1\leftarrow2}+\Gamma_{2\leftarrow1})^5}\ .
\end{split}
\label{eq_fcs}
\end{equation}
As pointed out by Jordan and Sukhorukov
\cite{jordan_2004,jordan_2005} these expressions are very general,
\mbox{\emph{i.\ e.}} they do not depend on the microscopic origin
of the rates or the current channels. For the 1-dot model the two
current channels were identified from phase space plots of the
oscillating dot as a \emph{shuttling} and a \emph{tunneling}
channel, respectively, with known analytic expressions for the
corresponding two currents \cite{flindt_2005, novotny_2004}. By
comparing the numerical results for the first two cumulants with
the corresponding analytic expressions given above, the two rates
$\Gamma_{1\leftarrow2}$ and $\Gamma_{2\leftarrow1}$ could be
extracted, and finally a comparison of the numerical results for
the third cumulant and the analytic expression given above (with
the extracted rates\footnote{In a certain limit the rates may even
be found analytically, see Ref. \cite{donarini_2004}.}
$\Gamma_{1\leftarrow2}$ and $\Gamma_{2\leftarrow1}$) confirmed the
conjecture about the bistable behavior (see Fig.
\ref{fig_1dotFCS}). This in turn explained a giant enhancement of
the zero-frequency current noise (the second cumulant) found in
Ref. \cite{novotny_2004}.

A similar enhancement of the zero-frequency current noise was
found in a study of the 3-dot model \cite{flindt_2004}. Also in
this case, the enhancement was tentatively attributed to a
switching behavior, however, neither the number nor the nature of
the individual current channels were clarified, and no
quantitative explanation could be given. Phase space plots of the
oscillating dot seem to indicate the existence of two current
channels \cite{flindt_2004}: One channel, where electrons tunnel
\emph{sequentially} through the array of dots, and one channel,
where electrons \emph{co-tunnel} between the static dots. The
current corresponding to each of the two channels can be read off
from the numerical results obtained in Ref. \cite{flindt_2004}. By
proceeding along the lines outlined above, the conjecture that the
enhanced noise is due to a slow switching between the sequential
and co-tunneling channel can be scrutinized.

\begin{figure}
  \includegraphics[height=.16\textheight, trim = 0 0 0 0, clip]{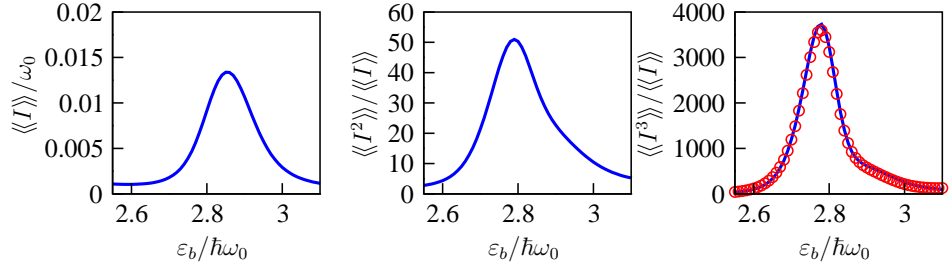}
  \caption{First three cumulants for the 3-dot model as a function of the device bias $\varepsilon_b$ (model parameters correspond to Fig. 4 in Ref. \cite{flindt_2004}). The sequential tunneling channel current is $I_{\mathrm{seq}}=0.043\omega_0$ and the co-tunneling channel current $I_{\mathrm{cot}}=0.0008\omega_0$ ($e=1$). Full lines
  indicate numerical results, while circles show the \mbox{(semi-)} analytic results for the third cumulant. The central panel shows the giant enhancement of the zero-frequency noise. (Left and central panel reproduced from Ref. \cite{flindt_2004}).}
  \label{fig_3dotFCS}
\end{figure}

\begin{figure}
  \includegraphics[height=0.16\textheight, trim = 0 0 0 0, clip]{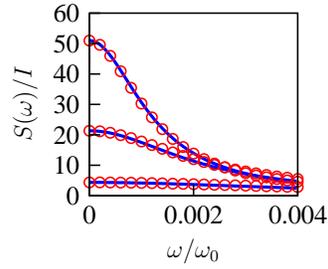}
  \caption{Finite-frequency current noise $S(\omega)$ (normalized with
  respect to the current $I$) for the 3-dot model. Circles
  indicate numerical results, while full lines are the
  corresponding (semi-)analytic results for a slow bistable switching process \cite{flindt_pre2005}. The results correspond to Fig. \ref{fig_3dotFCS} with $\varepsilon_b=2.60\hbar\omega_0$ (lower curve), $2.70\hbar\omega_0$, $2.79\hbar\omega_0$ (upper curve).}
  \label{fig_nonzero}
\end{figure}

\begin{figure}
  \includegraphics[height=.16\textheight, trim = 0 0 0 0, clip]{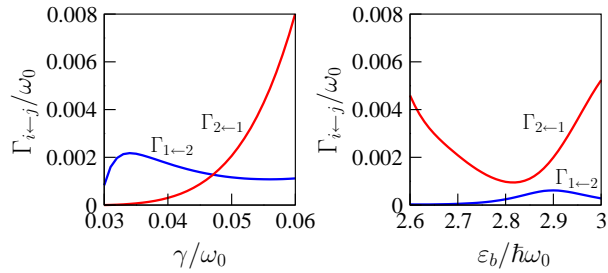}
  \caption{Left panel: The two switching rates for the 1-dot model as a function of the damping strength $\gamma$.  Here the two current channels are the shuttling channel (1) and the
  tunneling channel (2). The rates correspond to the results shown
  in Fig. \ref{fig_1dotFCS}.
  Right panel: The two switching rates for the 3-dot model as a function of the device bias $\varepsilon_b$. Here the two current channels are the sequential tunneling channel (1) and the
  co-tunneling channel (2). The rates correspond to the results shown
  in Fig. \ref{fig_3dotFCS}.}
  \label{fig_rates}
\end{figure}

\emph{Results} -- In Figs. \ref{fig_1dotFCS}, \ref{fig_3dotFCS} we
show numerical results for the first three cumulants for the two
models together with the analytic expression for the third
cumulant of a bistable system with rates extracted from the first
two cumulants. We take the agreement between the numerical and
(semi-) analytic results as evidence that both models exhibit a
bistable behavior. In Ref. \cite{flindt_pre2005} this
interpretation was further supported by numerical studies of the
finite-frequency current noise in the 1-dot model.
Correspondingly, we show in Fig. \ref{fig_nonzero} the agreement
between the numerical results for the finite-frequency noise in
the 3-dot model and semi-analytic results for a slow bistable
switching process \cite{flindt_pre2005}. In Fig. \ref{fig_rates}
we show the extracted rates for both models. Most noteworthy is
the crossing of the two rates in the 1-dot case, which results in
the change of sign of the third cumulant seen in Fig.
\ref{fig_1dotFCS}. On each side of the crossing one of the current
channels dominates. In the 3-dot case, the two rates close in,
however, without crossing each other. Consequently one of the
current channels, the sequential tunneling channel, never
dominates. It should also be noted that in both models one of the
currents is comparable to one of the rates, which implies that
some corrections to Eq. \ref{eq_fcs} are expected
\cite{jordan_2005}. However, we have found that these corrections
do not contribute significantly.

\emph{Conclusion} -- We have presented a study of noise in two
models of quantum shuttles. By evaluating numerically the first
three cumulants of the full counting statistics, we have shown
that a giant enhancement of the zero-frequency current noise in
both models can be explained in terms of a slow bistable switching
behavior. For both models, this interpretation is supported
further by a numerical evaluation of the finite-frequency current
noise. We underline that although the two models behave very
differently, it is the same mechanism that is responsible for the
giant enhancement of the noise.

\end{document}